Full length article

# Reliable machine learning potentials based on artificial neural network for graphene

Akash Singh[1], Yumeng Li [*,2]

*University of Illinois Urbana-Champaign, Department of Industrial and Enterprise Systems Engineering, 104 S Mathews Ave Urbana IL 61801, USA*

## ARTICLE INFO

## ABSTRACT

Graphene is one of the most researched two dimensional (2D) material in the past two decades due to its unique combination of mechanical, thermal and electrical properties. Special 2D structure of graphene enables it to exhibit a wide range of peculiar material properties like high Young's modulus, high specific strength, electrical conductivity etc. which are critical for myriad of applications including lightweight structural materials, multifunctional coating and flexible electronics. As it is quite challenging and costly to experimentally investigate graphene and graphene based nanocomposites, computational simulations such as molecular dynamics (MD) simulations are widely adopted for understanding the microscopic origins of their unique properties. However, disparate results were reported from computational studies, especially MD simulations using various empirical inter-atomic potentials. In this work, an artificial neural network (ANN) based interatomic force field potential has been developed for graphene to represent the potential energy surface based on first principle calculations. The developed machine learning potential (MLP) facilitates high fidelity MD simulations to approach the accuracy of ab initio methods but with a fraction of computational cost, which allows larger simulation size and length, and thereby enables accelerated discovery and design of graphene-based novel materials. Lattice parameter, coefficient of thermal expansion (CTE), Young's modulus and yield strength are estimated using machine learning accelerated MD simulations (MLMD), which are compared to experimental and first principle calculations from previous literatures. It is demonstrated that MLMD can capture the dominating mechanism governing the CTE of graphene, including effects from lattice parameter and out of plane rippling. The MLMD approach is highly scalable for 2D materials and can help in accelerating the research of novel 2D materials and 2D material hybrids with unique atomic structures.

## 1. Introduction

2D materials exhibit novel properties for a myriad of advanced applications like lightweight materials, energy storage, and flexible electronics. After the discovery of graphene scientific community quickly discovered other 2D materials such as hBS, Phosphorene, $MoS_2$, $MoSe_2$ [1,2] etc. Graphene is a unique 2D material due to its excellent mechanical, electrical and chemical properties [3,4] and thus has received widespread attention of scientific community in the past two decades. The 2D structure of carbon atoms in graphene makes it one of the lightest and strongest material present on this planet [5]. The $sp^2$ hybridization (covalent bonds) of carbon in graphene provides it a stable 2D structure resulting in a material with a very high surface area to volume ratio. These covalent bonds provide graphene a high in-plane tensile strength. Out of four valence electrons in graphene three electrons form covalent bonds with other carbon atoms leaving one excess electron in valence shell providing electrical and thermal conductivity [6,7] to graphene. These unique thermal, electrical and mechanical properties makes graphene an excellent candidate for creating composite materials. Thus, graphene finds itself useful in diverse areas and has several applications in mechanical resonators, batteries, gas detectors etc. So, graphene can be touted as a metamaterial with endless possibilities which can revolutionize the future of composites, batteries, superconductors [6] etc.

In the past decade, graphene and graphene based nanocomposites have been studied experimentally to understand its unique material behavior and develop its structure–property relationships. Lee et al. [8] studied Young's modulus of graphene and estimated it to be 1 ± 0.1 TPa. His studies also established that graphene was the strongest material ever measured, till date. Another study by Lee U. et al. [9]

---

* Corresponding author.
  *E-mail addresses:* akashs5@illinois.edu (A. Singh), yumengl@illinois.edu (Y. Li).
[1] Graduate Student.
[2] Assistant Professor.

using Raman spectrography evaluated Young's modulus for single layer graphene to be 2.4 TPa. Bao et al. [10] evaluated coefficient of thermal expansion of graphene experimentally and found it to be negative at room temperature which transitions to positive at 375 K while increasing monotonically from 300 K to 400 K. Another experimental study conducted by Yoon et al. [11] evaluated the coefficient of thermal expansion using Raman spectrography. Yoon et al. [11] predicted coefficient of thermal expansion to be monotonically increasing but remains negative throughout the temperature range of 200 K to 400 K. Yang et al. [12] compiled data from previous experimental studies for in-plane lattice parameter of graphene. His studies suggested a constant lattice parameter of 2.46 Å within a temperature range of 200 K–1000 K. Manigandan et al. [13] studied topological properties of graphene based nanocomposites doped in Kevlar. His study suggested an increase in nanocomposite strength with increase in the graphene's percentage in nanocomposite from 0%–5%. Discrepant data have been reported in the experimental studies of graphene for Young's modulus and coefficient of thermal expansion. This could be attributed to extremely small thickness of 2D materials, which makes experimental characterization and testing of graphene challenging and costly. Therefore it is compelling to seek alternative methods to systematically evaluate material properties of graphene.

Analytical and computational simulations, as compared to experimental studies are usually cost and time efficient methods to investigate 2D materials. First principle calculations and classical MD simulations are widely adopted for the study of 2D materials [14,15]. First principle calculations using Density Functional Theory (DFT) simulations and ab intio MD simulations investigates electronic structure of many body systems to evaluate material properties. Tianjiao et al. [14] used first principle calculations with quasi harmonic approximations (QHA) to evaluate Young's modulus of graphene over a temperature range of 20 K to 1000 K to have a constant value of 1.15 Tpa. Mounet et al. [15] used DFT simulations to evaluate lattice parameter of graphene and found it to be slightly decreasing with temperature in range of 20 K to 1000 K. His study also evaluated coefficient of thermal expansion of graphene (CTE) with respect to temperature. CTE was found to be negative with a non-monotonic shape which decreases from 20 K to 250 K and then increases from 250 K to 2000 K. Classical MD simulations are also widely used to establish, evaluate and analyze material properties because of its computational efficiency [16–18]. These simulations adopt simplified empirical potentials which are usually derived from parameter fitting based on either experimental or first principle calculations results. Thus, the success of classical MD simulations is highly determined by the fidelity and availability of empirical potentials. Using classical MD simulations, Rahman et al. [19] investigated mechanical performance and fracture behavior of silicon doped graphene. In another work using classical MD simulations, Li et al. [18] established the effect of functionalized interface between graphene - polyethylene and established that this particular composite exhibits different Young's modulus with different number of functional groups. These studies [18,19] show that mechanical properties of graphene and graphene based nanocomposites can be easily characterized by classical MD simulations which otherwise was difficult to establish experimentally. It also proves the ability of classical MD simulations to simulate larger/realistic material models with reasonable accuracy that is currently difficult for DFT simulations and ab initio MD simulations.

First principle calculation includes electronic structure of atoms to evaluate material property thereby providing highly accurate results. But this accuracy comes at a computational cost which limits the simulation size and simulation time for the material system being investigated. Thus, we find ourselves in a stalemate situation where one type of simulation is fast (classical MD) and the other one is accurate (DFT). In the recent years, a few studies have been conducted for representing the potential energy surface of materials derived from DFT simulations using machine learning techniques [20,21]. In this approach, a machine learning model is trained to represent the potential energy surface for a given material system which is highly scalable and transformable. The trained machine learning model thus can be used in classical MD simulations to understand material behaviors and evaluate material properties. Such machine learning based interatomic potential enables us to combine the accuracy of DFT simulations with the computational efficiency of classical MD simulations [20,22].

In the present study, an artificial neural network based machine learning interatomic potential (MLP) was developed for graphene. Training data was generated from first principle calculations of graphene following the work of Rowe et al. [22]. Atomic structures in the training data were encoded by an atom centered coordinate system using symmetry functions. The trained ANN with optimal weights and biases represents the interatomic potentials of graphene under various loading and experimental conditions. LAMMPS [23], an open source MD simulation software was used to conduct classical MD simulations in this study. Another open source software n2p2 [24] was used with LAMMPS to incorporate machine learning interatomic potentials. In this study, lattice parameter, coefficient of thermal expansion, Young's modulus and ultimate tensile strength, are evaluated to validate the ANN-based MLP. Finally, comparison of these material properties of graphene with experimental and DFT simulations demonstrates the effectiveness and practical use case of machine learning interatomic potentials for graphene.

## 2. Method

In this study, ANN based machine learning interatomic potentials were developed to establish the potential energy surface for graphene. In order to prepare suitable training data for ANN, all the atomic structures were represented using symmetry functions [25], which is an invariant representation of atom's coordinates in the material system. During the training process, weights and biases of ANN were optimized at each iteration to reduce the error between predicted energy and reference energy via backpropagation. Training iterations were stopped for a given ANN once a desired level of accuracy was achieved for ANN's weights and biases. This set of weights and biases for a trained ANN creates a complex function which defines the structure–energy relationship and is referred to as machine learning interatomic potential for graphene. This section is organized into two subsections. Section 2.1 discusses the symmetry functions used for encoding atomic coordinates and Section 2.2 discusses training methodology for ANN.

### 2.1. Symmetry functions

Before training an ANN, each atomic structure in the training dataset needs to be described by an invariant representation which is independent of translation, rotation and exchange of equivalent atoms. In this study, this was achieved by encoding the atomic structures using symmetry functions [25,26]. Using this approach each reference atomic structure was uniformly represented which eliminated the need of coordinate system. Since, the basis functions were atom centered thus, the energy evaluation after training of machine learning interatomic potentials remains independent to the choice of coordinate system. This eliminated the need of multiple coordinate transformations and thus, is an optimal method to evaluate the energy of datasets with multiple coordinate systems.

Two type of symmetry functions were adopted for fingerprinting the atomic structures in this study i.e. radial symmetry functions and angular symmetry functions. Radial symmetry functions, as shown in Eqs. (1) and (2), consider two-body interactions around the center atom while the angular symmetry functions, as shown in Eq. (3), consider three-body interactions around the center atom. Radial and angular symmetry functions in conjunction gives a complete description of the atomic environment of a material system with one atomic species. A radial symmetry functions is typically described as follows:

$$G_r^i(\sigma) = \sum_{j=i}^{nr} g_r^i(R_{ij}) = \sum_{j=i}^{nr} \exp^{-\eta(R_{ij}-R_s)^2} f_c(R_{ij}) \tag{1}$$

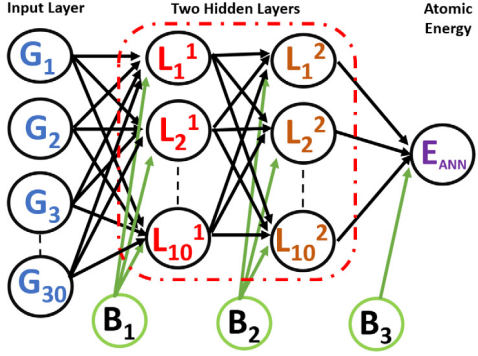

**Fig. 1.** Schematic of atomic artificial neural network used in this study. Input to the atomic ANN is symmetry function values. ANN has two hidden layers with 10 neurons each. Output from the atomic ANN is atomic energy of the given atom.

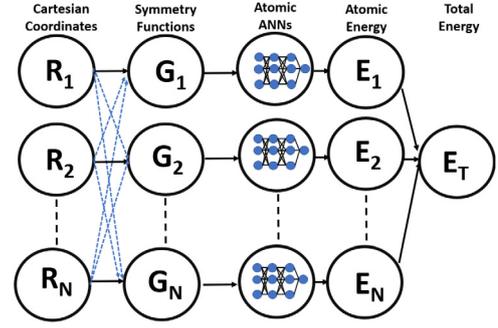

**Fig. 2.** Schematic of neural network potential for graphene. Each atom is represented by a line with cartesian coordinates $R_i$. First step converts cartesian coordinates into symmetry function values $G_i$ depending on the cartesian coordinate of all the other atoms in the local environment, shown by blue dotted arrows. Symmetry function values is fed to ANN providing energy contribution from each atom $E_i$. Finally the total structural energy is evaluated as the sum of all $E_i$.

$$f_c(R_{ij}) = \begin{cases} 0.5[cos(\frac{\pi R_{ij}}{R_c}) + 1], & \text{for } R_{ij} \leq R_c \\ 0, & \text{otherwise} \end{cases} \quad (2)$$

where $\sigma$ represents a three dimensional atomic coordinates, $R_{ij} = |R_i - R_j|$ is the radial distance between a central atom $i$ and its neighboring atom $j$, $R_c$ is the cutoff radius, $\eta$ is the radial parameter, $nr$ is the number of neighboring atom around atom $i$ within a cutoff distance and $f_c$ is the cutoff function. An angular symmetry function is typically described as follows:

$$G_a^i(\theta) = \sum_{j=i}^{na} \sum_{k} g_a^i(\theta_{ijk})$$
$$= 2^{1-\zeta} \sum_{j \neq i} \sum_{k=j}^{na} \sum_{=i} (1 + \lambda cos(\theta_{ijk}))^\zeta \exp^{-\eta(R_{ij}^2 + R_{ik}^2 + R_{kj}^2)} f_c \quad (3)$$

where $\theta_{ijk}$ is the angle between central atom $i$ and its neighboring atoms $j$ and $k$. $\zeta$, $\lambda$ and $\eta$ are angular parameters used to define the symmetry function. $na$ is number of atoms around atom $i$ within cutoff distance and $f_c$ is the cutoff function.

A vector $G_i = [G_{r1}, G_{r2}, \ldots, G_{r10}, G_{a1}, G_{a2}, \ldots, G_{a20}]$ based on the set of radial and angular symmetry functions was used to encode the atomic environment for each atom in the reference dataset. Now, the total energy of a single structure can be written as sum of energy from individual atom which can be further written in terms of radial and angular symmetry functions, as shown in Eq. (4). Total energy of a single structure can be described as follows:

$$E_{tot} = \sum_{i=1}^{n} E_i \approx \sum_{i=1}^{n} E_i^{ANN}(G_r(R), G_a(R, \theta)) \quad (4)$$

where $E_{tot}$ is the total energy of a reference dataset, $E_i$ is the energy of individual atom $i$, $n$ is the total number of atoms in the structure, $E_i^{ANN}(G_r(R), G_a(R, \theta))$ represents energy of atom $i$ from artificial neural network which is a function of radial and angular symmetry functions. Thus, for each structure the input to artificial neural network is a matrix of $n$ by $m$ where $n$ is the number of atom in the structure and $m$ is the total number of symmetry functions for each atom.

### 2.2. Artificial neural network

Feedforward artificial neural network [27] is one of earliest neural network which was developed in the field of machine learning. In this neural network information moves in only one direction and does not cycle, as shown in Fig. 1. The structure of an ANN is defined by 3 parameters i.e number of hidden layers, activation functions and number of nodes in the hidden layers.

An energy-based loss function was adopted for ANN training which minimized the error between predicted energy from ANN and reference energy from dataset generated using DFT simulations. The root mean squared error (RMSE) function was used as the loss function in this study as shown in Eq. (5).

$$\epsilon(w_m, b_m) = \frac{1}{2} \sum_{s=1}^{c} [E_s^{ANN}(\sigma, w, b) - E_s^{ref}]^2$$
$$= \frac{1}{2} \sum_{s=1}^{c} |\sum_{i=1}^{n} E_{s,i}^{ANN}((G_r^i G_a^i, w, b)) - E_s^{ref}|^2 \quad (5)$$

where $c$ is the number of configurations in the dataset and $ref$ is used to denote the reference dataset, $E_s^{ANN}$ is the energy obtained from ANN and $E_s^{ref}$ is the reference energy evaluated using DFT calculations. Since, ANN predicts energy of individual atoms while no atomic energies can be provided by DFT calculations, therefore atomic ANNs are trained simultaneously over all atoms in each dataset and summed over to get the total energy, as shown in Fig. 2.

During the training of atomic ANNs, the gradient of loss function with respect to weight parameters was estimated using back propagation method to facilitate iterative weight updates. Two different training methods, Gradient Descent (GD) and Levenberg–Marquardt (LM), were used to optimize the ANN weights. The weight parameters were then updated using the following functions in each training epoch.

$$w^{i+1} = w^i + \Delta w^i \quad (6)$$

For Gradient Descent

$$\Delta w^i = -h \nabla \epsilon \quad (7)$$

For Levenberg–Marquardt

$$\Delta w^i = -(J^{T,i-1} J^{i-1} + \lambda I)^{-1} \nabla \epsilon \quad (8)$$

$$\nabla \epsilon = \frac{\delta \frac{1}{2} \sum_{s=1}^{c} [E_s^{ANN}(\sigma, w, b) - E_s^{ref}]^2}{\delta w_m}$$
$$= \sum_{s=1}^{c} [E_s^{ANN} - E_s^{ref}]^2 \sum_{i=1}^{n} \frac{\delta [E_{s,i}^{ANN}(G_r^i, G_a^i, w, b)]}{\delta w_m} \quad (9)$$

As compared to Gradient Descent, Levenberg-Marquardt method converges quickly due to its second order correction. The dimension of Hessian matrix that needs to be inverted quickly escalates if the dimension of artificial neural network becomes too large. Then, we need to move to some other training method like BFGS but fortunately our training dataset was not that complicated.

## 3. Machine learning potentials for graphene

This section details the development of machine learning interatomic potential for graphene. Section 3.1 introduces the generation of training dataset, Section 3.2 discusses the graphene's atomic environment description, Section 3.3 discusses two training methods for MLP and their effectiveness and Section 3.4 illustrates the validation steps followed to evaluate the performance of MLP.

### 3.1. Training dataset

Due to the regressive nature of machine learning, the quality of developed MLP is largely determined by representative structures in the training dataset. Our training datasets were generated based on tightly converged DFT simulations conducted for reference configurations sampled from various MD simulation trajectories. For the current work, all the reference structures of graphene contains one single layer with 24 carbon atoms. MD simulations were used to stretch graphene sheets and generate different atomic structures which were then imported to DFT to get reference energy for training. A total of 320 reference configurations were collected by capturing evenly from stretched trajectories of graphene i.e. 160 atomic structural snapshots were captured by stretching graphene in armchair direction and the remaining 160 atomic structural snapshots were captured by stretching graphene in the zigzag direction.

The total potential energy of each reference configuration was calculated with DFT simulations using PBE exchange–correlation functional as implemented in PWSCF of Quantum ESPRESSO package [28]. Wave functions were represented by in-plane wave basis sets with energy cutoffs of 40 Ry and 120 Ry, respectively, using GBRV ultrasoft pseudopotentials for the core regions of atom. For the Brillouin zone integration, all calculations employed gamma centered k-point meshes (4-16-1 k-points) for the crystal cell with 24 carbon atoms.

### 3.2. Graphene atomic environment description

Graphene is relatively a simple material system containing only carbon atoms. In general, symmetry function parameters are selected to increase the stability and fidelity of developed MLP. Artirth's strategy [29] for the selection of descriptors for symmetry functions was adopted in this study. In this strategy combined descriptors were used to represent local structure and composition. It was shown by Artirth [29] that non linear machine learning models did not required a mathematically complete descriptor set as long as it was able to differentiate between all relevant features. Thus, the potential energy surface can be plotted with a small basis set which corresponds to a coarse representation of radial and angular symmetry functions.

Radial symmetry function, as shown in Eq. (2), is a combination of Gaussian function and a cutoff function. The cutoff function $f_c$ ensures that only energetically relevant regions close to the center atom are encoded in atom centered symmetry functions. $\eta$ and $R_s$ modulates the width and position of the Gaussian function respectively. For angular symmetry function, as shown in Eq. (3), the term in bracket characterizes the distribution of angles. Parameter $\lambda$ shifts the maximum value of angular term between 0° and 180° whereas parameter $\zeta$ controls the width of angular Gaussian function. The introduction of terms based on $r_{ij}$ introduces asymmetric behavior into angular functions. Thus, a total of four parameters ($\eta, R_s, \lambda, \zeta$) needs to be selected to form suitable sets of atom centered symmetry functions. The choice of $\lambda$ and $\zeta$ for angular symmetry functions is relatively straightforward. In general, using two sets of angular symmetry functions with $\lambda = 1$ and $\lambda = -1$, respectively are used as it covers all possible ranges of angles present in the environment. $\zeta$ dictates the width of angular resolution. With increasing values of $\zeta$ less angular resolution can be achieved thereby making $\zeta = 2$ to be more important than the higher values of $\zeta$. In our symmetry functions we have taken values of $\zeta$ from 2 to 24 in increments of 2 units. $R_s$ shifts the Gaussian function laterally. For all the symmetry functions used in this study $R_s = 0$ has been used. $\eta$ has the largest influence on descriptor performance for radial as well as angular symmetry functions. Usually the spatial extension of radial symmetry function with smallest effective range (shortest interatomic distances present in the data set) dictates the highest value of $\eta$. And the lowest value of $\eta$ in radial symmetry functions is guided by correlation between the values of a given symmetry function (at least 90% suggesting linear independence) for all atoms in the reference set. In order to achieve a balanced coverage of space from $r_o$ to $r_c$ an auxiliary radial grid is introduced. This auxiliary grid consists of $N$ equally spaced points ranging from $r_o$ to $r_c$. Thus, $\eta_i = 1/2r_i^2$ defines the values of $\eta$ with the above constraints. Finally in our study, 10 radial and 20 angular symmetry functions were used to describe the local neighboring atomic environment in the simulation system, i.e. a 30-dimensional input vector for the atomic ANN training. All the parameters used for calculating radial and angular symmetry functions are listed in Tables 1 and 2.

**Table 1**
Parameters for $G_r$ type of radial symmetry functions ($R_c = 6.5$ Å and $R_s = 0$).

| S.No | $\eta$ | S.No | $\eta$ |
|---|---|---|---|
| 1 | 0.004938 | 6 | 0.493827 |
| 2 | 0.012404 | 7 | 1.240438 |
| 3 | 0.031158 | 8 | 3.115839 |
| 4 | 0.078266 | 9 | 7.826633 |
| 5 | 0.196596 | 10 | 19.659613 |

**Table 2**
Parameters for $G_a$ type of angular symmetry functions ($R_c = 6.5$ Å and $R_s = 0$).

| S.No | $\eta$ | $\zeta$ | $\lambda$ | S.No | $\eta$ | $\zeta$ | $\lambda$ |
|---|---|---|---|---|---|---|---|
| 1 | 0.000049 | 2 | 1 | 11 | 0.000049 | 2 | -1 |
| 2 | 0.000124 | 4 | 1 | 12 | 0.000124 | 4 | -1 |
| 3 | 0.000312 | 6 | 1 | 13 | 0.000312 | 6 | -1 |
| 4 | 0.000783 | 8 | 1 | 14 | 0.000783 | 8 | -1 |
| 5 | 0.001966 | 10 | 1 | 15 | 0.001966 | 10 | -1 |
| 6 | 0.004938 | 12 | 1 | 16 | 0.004938 | 12 | -1 |
| 7 | 0.012404 | 14 | 1 | 17 | 0.012404 | 14 | -1 |
| 8 | 0.031158 | 16 | 1 | 18 | 0.031158 | 16 | -1 |
| 9 | 0.078266 | 18 | 1 | 19 | 0.078266 | 18 | -1 |
| 10 | 0.196596 | 20 | 1 | 20 | 0.196596 | 20 | -1 |

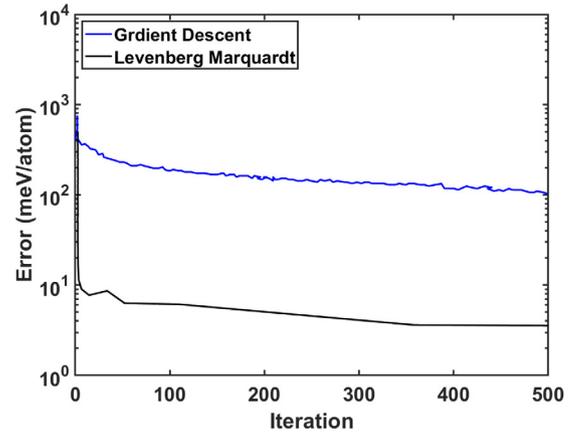

**Fig. 3.** Training error of MLP with error history using gradient descent and Levenberg–Marquardt backpropagation algorithm.

### 3.3. MLP training

This study uses an atom-based high-dimensional neural net as shown in Fig. 2 consisting $N$ atoms where each line represents one atom of graphene. Firstly, cartesian coordinates $R_i = (X_i, Y_i, Z_i)$ are transformed into symmetry functions $G_i$ which act as suitable input for atomic ANNs. These atomic ANNs then predicts the energy for each atom which is eventually summed up get total energy of the structure. The structure of feedforward ANN used in this study had 2 hidden layers with 10 nodes in each hidden layer. Hyperbolic tangent function was used as the activation function in the hidden layers for training of MLP. The reference dataset was divided to keep 90% of dataset for training MLP and the remaining 10% for testing of MLP. The training progress with optimization iterations was typically monitored through the evolution of the root mean squared error (RMSE) of the reference structures, as shown in Eq. (10).

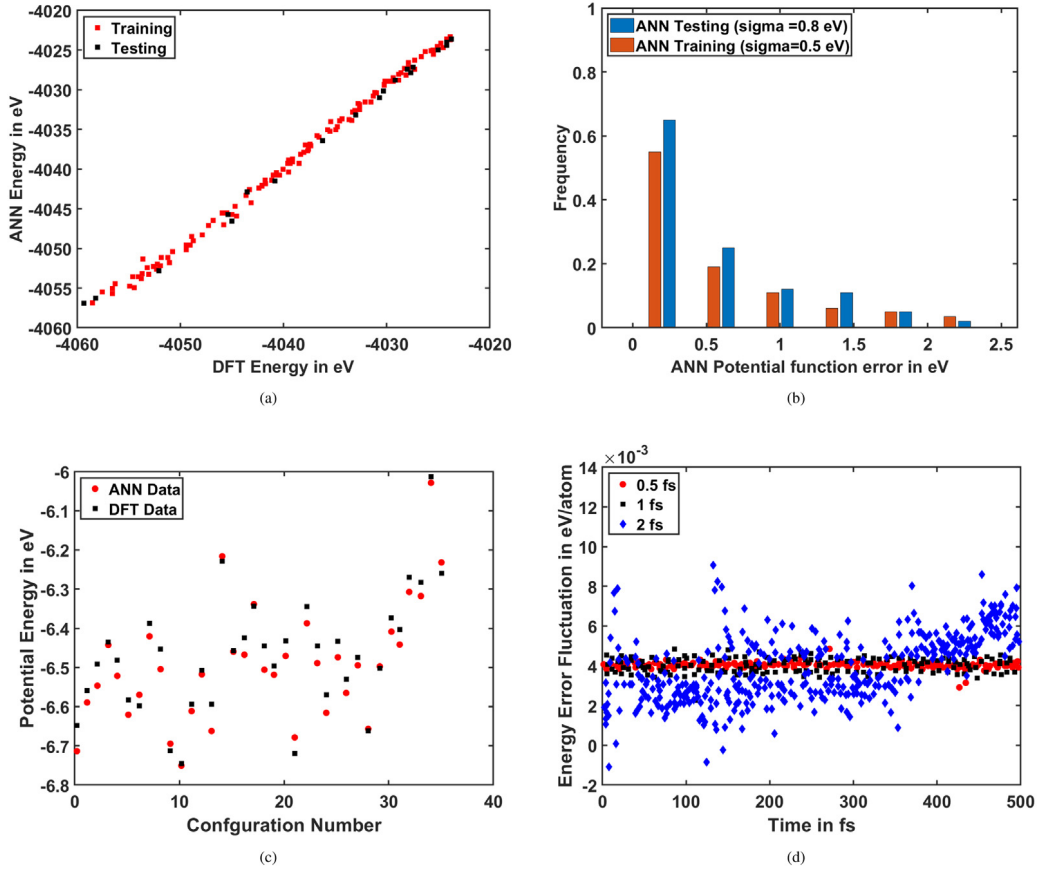

**Fig. 4.** (a) Correlation of ANN energies with training and testing data set with DFT energies (b) Histogram of energy error distribution for training and testing datasets (c) Potential energy fluctuations for different configurations for ANN and DFT simulations (d) Effect of different time steps on potential energy.

$$RMSE = \sqrt{\frac{1}{n} \sum_{i=1}^{nt} (E_i - E_i^{ANN})^2} \tag{10}$$

Two training methods, Gradient Descent (GD) and Levenberg–Marquardt (LM), were implemented and compared in this study. It is known that GD method is computationally less demanding as compared to LM method. For comparing these two methods error histories of predicted energy with respect to the reference DFT values during the training process is plotted, as shown in Fig. 3. It can be noted that both training methods can achieve well converged ANN structures within reasonable number of iterations. However, it can be seen that LM method can reach an accuracy of 5 meV/atom in less than 50 iterations while the GD method converges to a much lower accuracy with an error of 115 meV/atom after 400 iterations. Although the number of training iterations does not directly translate to the computing time. Considering the dimension of current ANN and relatively small training dataset used in current work, LM method was found to be more efficient. Thus, weights and biases corresponding to LM training method were used to develop the current version of MLP.

### 3.4. Validation of machine learning potential

MLP validation is an important step to verify the functionality of interatomic potentials and its prediction power. Fig. 4(a) shows the correlation between predicted MLP energies and corresponding DFT reference energies for atomic configurations in training and testing datasets. It can be noted that for both training and testing datasets predicted energy and reference energy shows good correlation as indicated by fitting function $y = x$. This implies that energies of atomic structures throughout the interested phase space was accurately captured by MLP. Fig. 4(b) shows histogram of energy error distribution for testing and training datasets in electron-Volts. It can be found that about 75% of training error was less than the target accuracy while about 70% of testing error was less than the target accuracy. Another test to assess the reliability of MLP was monitoring the potential energy fluctuation in different MD simulation trajectories. Fig. 4(c) shows ANN energies compared with DFT energies of 36 reference structures. It can be found that MLP performs very well in predicting energy of metastable atomic structures with deviations smaller than 3 meV/atom. It is also essential for the MLP to have a smooth and continuous potential energy surface to enable the numerical integration of equations of motion in MD simulations. Fig. 4(d) shows the total energy during MD simulation of a single layer graphene sheet over 500 fs using time steps of 0.5 fs, 1 fs and 2 fs with NVE ensemble at room temperature. We can observe that total energy of the material system was well conserved and fluctuations were constrained within few meV for MD simulations using the step size of 0.5 fs and 1 fs. However, a small drift was observed for the case using step size of 2 fs, which needs further investigation.

Another important metric to evaluate the quality of MLP can be the accuracy with which atomic forces can be predicted by MLP. As our MLP was trained only based on potential energies from DFT simulations it was critical to investigate its capability on force prediction. 300 random samples from MD simulations were chosen to compare forces as predicted by MLP with DFT simulation results. As shown in Fig. 5 some scattering can be observed but forces predicted by MLP strongly correlated with DFT results and were within 1.5 eV/Å of standard deviation. In addition, Fig. 6 shows the direction of majority of force vectors predicted by MLP which were found to be within 20 degrees of DFT simulation results.

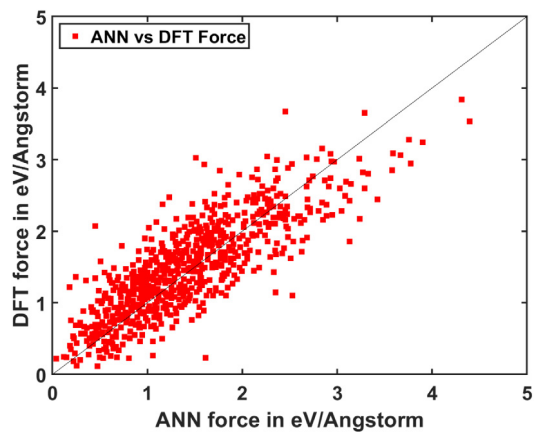

**Fig. 5.** Correlation of magnitude of atomic forces in graphene predicted by ANN and DFT simulations.

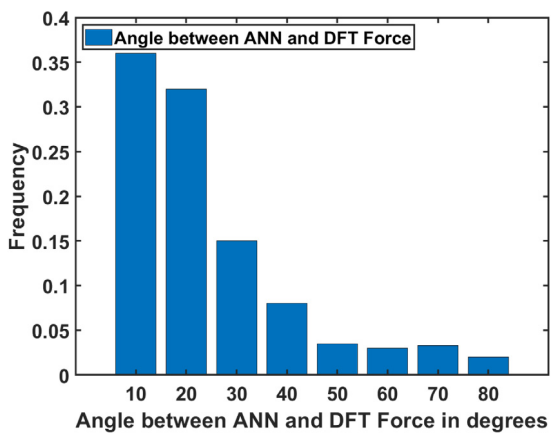

**Fig. 6.** Histogram plot showing similarity in angle between predicted ANN and DFT forces.

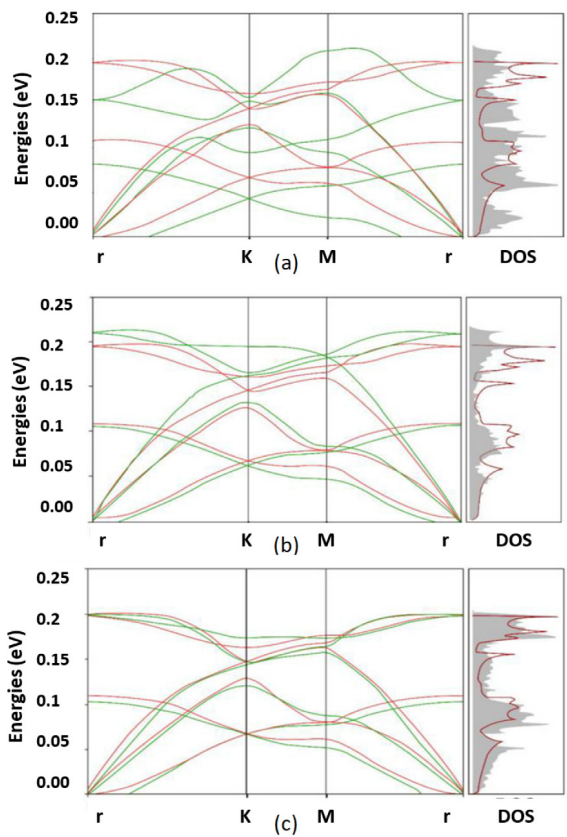

**Fig. 7.** Comparison of predicted phonon dispersion using MLP (small displacement method) with experimental results (a) using 124 samples, cutoff distance 4.5 Å (b) using 300 samples, cutoff distance 4.5 Å (c) using 300 samples with increased cutoff distance of 6.5 Å.

In atomistic simulations, calibration of phonon dispersion is a fundamental requirement of interatomic potentials to have a correct description of lattice dynamics. Phonon dispersion is an experimentally measurable property of a material which also serves as an excellent independent metric to validate the overall quality of an inter-atomic potential. Therefore, validation of phonon dispersion was conducted for MLP using small displacement method. Fig. 7 shows the phonon map of graphene predicted using MLP with different reference dataset. Green lines shows the results of atomistic simulations and red lines shows the results of experimental studies. As shown in Fig. 7a, an MLP trained with 124 sample configurations correctly predicted the shape and trend of most of the phonon branches but underestimated the absolute value. In Fig. 7b, by increasing sample size to 300, an evident improvement in the predicted phonon energies can be observed with highly correlated predictions for low energy branches while overestimations were still observed for high energy branches. By varying the representation strategy of atomic configurations in the reference dataset (i.e. increasing cutoff distance from 4.5 Å to 6.5 Å), better predictions for phonon dispersion were observed, as shown in Fig. 7c.

This indicates that quality of MLP can be systematically improved by tuning several parameters including size of training dataset and level of "digitization" process. Further systematic uncertainty quantification is required to identify a complete set of factors determining an optimal tuning strategy for MLP.

## 4. Prediction of thermal and mechanical properties

This section presents temperature dependence of mechanical and thermal properties of graphene by conducting MLMD simulations. Section 4.1 discusses the evaluation of thermal properties of graphene i.e. coefficient of thermal expansion and lattice parameter in temperature range of 125 K–1000 K. Section 4.2 discusses the mechanical performance of graphene i.e. Young's modulus and ultimate tensile strength in temperature range of 125 K–1000 K. These estimated mechanical and thermal properties of graphene are also compared with first principle calculation and experimental results from previous literatures.

### 4.1. Lattice parameter and coefficient of thermal expansion

Lattice parameter is one of the most fundamental property that can be predicted for an atomistic model of graphene. Lattice parameter can affect many intrinsic properties of graphene such as Young's modulus, yield strength, coefficient of thermal expansion etc. In a nanocomposite, the type and degree of interaction between graphene and polymers can lead to high variation in material properties depending on the degree of lattice matching between two materials [30]. We compare lattice parameters predicted by MLMD simulations with predictions from computational and experimental studies in the literature [14,15,31–33] as shown in Figs. 8(a) and 8(c). In-plane lattice parameter was evaluated for freestanding graphene sheet containing 12,400 atoms between temperature range of 125 K and 1000 K. MD simulations were performed with NPT ensemble using (Nosé–Hoover thermostat) isobaric–isothermal conditions with pressure set at 1 bar. Simulations at 300 K were equilibrated for 10 ns and three realizations were observed conducted for each temperature. The lattice parameters were calculated and averaged over a 20 ns duration for each simulation.

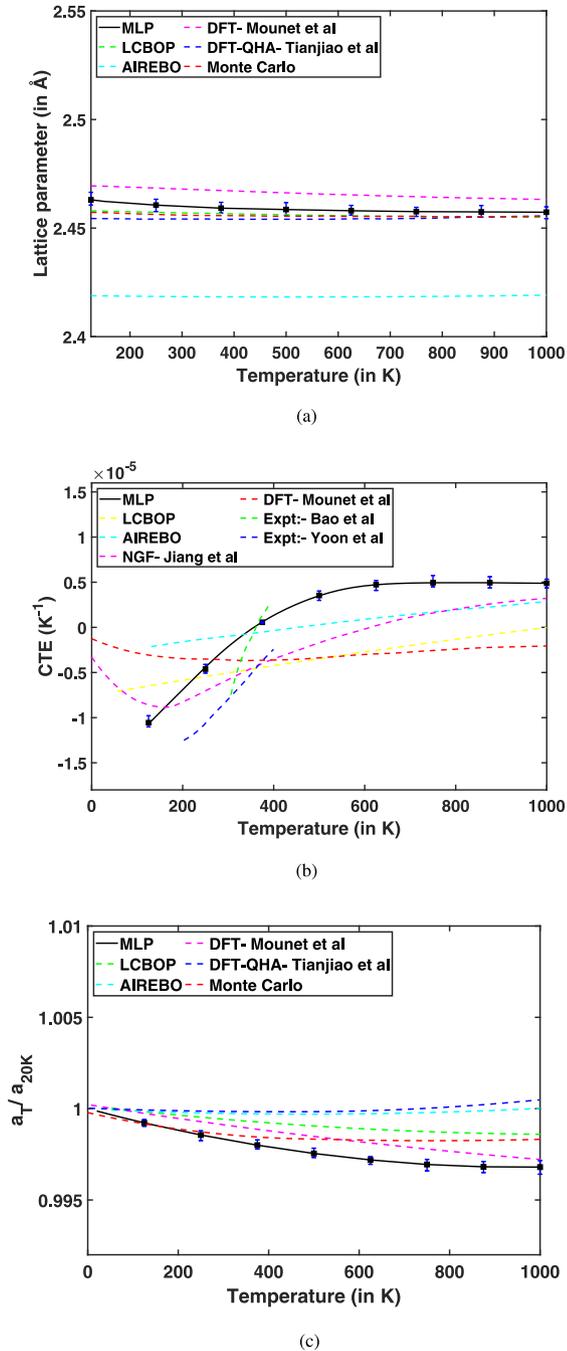

**Fig. 8.** (a) Temperature dependence of lattice parameter for different inter-atomic potentials and their comparison with machine learning potentials (MLP) (b) Computationally evaluated coefficient of thermal expansion for graphene as a function of temperature between 125 K to 1000 K for different inter-atomic potentials and their comparison with MLP (c) Thermal dependence of normalized lattice parameter with respect to the predicted values at 20 K - a varied range of predictions were observed i.e. monotonically increasing, decreasing to more complex non monotonic behavior. Third order polynomial fitting was used to evaluate lattice parameter value at 20 K.

Fig. 8(a) indicates that DFT simulations by Mounet et al. [15] predicts a slightly decreasing lattice parameter with 2.469 Å at 125 K and 2.463 Å at 1000 K. Another DFT result based on quasi harmonic approximation (QHA) by Tianjiao et al. [14] predicts a constant lattice parameter around 2.454 Å over the temperature range of 125 K to 1000 K. Monte Carlo simulations by Zakharchenko et al. [34] and classic MD simulations using empirical potential LCBOP [31] suggest a similar prediction of lattice parameter which stays constant at 2.457 Å over the temperature range of 125 K–1000 K. Classic MD simulation using AIREBO [32] potential predicts a much smaller lattice parameter 2.418 Å relative to the predictions in other works. Our MLMD simulation predicts a comparable lattice parameter with DFT simulations (e.g. around 2.456 Å) within the investigated temperature range. Fig. 8(c) shows normalized lattice parameter with respect to the value of lattice parameter at 20 K to manifest the change of lattice parameter over temperature. DFT simulations from Mounet et al. shows a monotonically decreasing lattice parameter, while DFT-QHA simulations by Tianjiao et al. shows a non-monotonic and slightly increasing lattice parameter in temperature range of 125–1000 K. Classic MD simulations based on AIREBO potential shows a similar non monotonic trend with DFT -QHA simulation results, while MD simulations using LCBOP potential shows a monotonic decreasing lattice parameter. Our MLMD simulation results follow a similar trend as the DFT simulation by Mounet et al. [15] indicating a monotonically decreasing lattice parameter between 20–1000 K.

Coefficient of thermal expansion (CTE) is another important thermal property of graphene as it provides insights into the anharmonicity of bonding interactions, relative strengths of in-plane and out-of-plane forces as well as vibrational coupling between harmonic and anharmonic modes. Eq. (11) shows the mathematical formulation used in this study for calculating coefficient of thermal expansion for graphene.

$$CTE = \frac{1}{A_t} \frac{\partial A_t}{\partial T} \qquad (11)$$

where $A_t$ is area of graphene sheet at temperature $T$ in Kelvin. For evaluating derivative of area with respect to temperature, spline interpolation was used between evaluated area at each temperature. In-plane CTE was computed using MLMD simulations on freestanding graphene sheets containing 12,400 atoms between a temperature range of 125 K and 1000 K. CTE was evaluated at discrete temperatures within this temperature range. MD simulations were performed with NPT ensemble using (Nosé–Hoover thermostat) isobaric–isothermal conditions with pressure set at 1 bar and temperature set at 300 K. Simulations were equilibrated for 10 ns at 300 K followed by a temperature increase to reach the desired temperature with a ramping rate of 25 K/ns. Three similar simulations were conducted for CTE at a particular temperature and final CTE was averaged out for each temperature.

The behavior of CTE with respect to temperature for graphene is an elusive topic where conflicting results have been found in the past. Experimental study of Bao et al. [10] suggests a transition of CTE from negative to positive (375 K) in temperature range of 300–400 K whereas another experimental study by Yoon et al. [11] predicts only negative CTE in temperature range of 200–400 K. Overall experimental studies suggest that coefficient of thermal expansion of graphene is negative at moderate temperatures (0–500 K) [10,11] i.e low lying bending phonon modes cause graphene to crumple and thereby shrink in in-plane direction. Raman spectroscopy and micromechanical measurements suggests graphene to have a negative in-plane coefficient of thermal expansion between temperature range of 30 K–300 K [35–37]. However, in a typical experimental evaluation of graphene has graphene adsorbed on a substrate material, thus an induced strain from substrate significantly affects lattice parameter and coefficient of thermal expansion value, thus leaving the study of freestanding graphene to be an eye-catching topic for theoretical scientists [34, 38]. Ab-initio MD simulation results broadly agree that coefficient of thermal expansion of graphene is negative over moderate temperature range but differ in predicting its magnitude. Non equilibrium green function (NGF) [39] study shows a non monotonic behavior of CTE with a negative CTE observed at moderate temperatures and positive CTE is observed at high temperatures (500 K and above). CTE switches sign at 600 K in this study as can be seen from Fig. 8(b). Another study using density functional theory by Mounet et al. [15] also predicts a non-monotonic behavior for CTE, but CTE remains negative up to 1000 K, with a minimum value observed at 300 K. Studies using empirical potential shows even more variations in predicting CTE. Adaptive intermolecular reactive empirical bond order (AIREBO) potentials predicts monotonously increasing CTE which remains positive over wide temperature range and only going negative below 200 K. Another empirical potential, long-range bond-order potential (LCBOP) predicts CTE to be entirely negative between 0–1000 K but with a monotonously increasing behavior [40].

Our MLMD simulations, as shown in Fig. 8(b), predict a similar trend of CTE changing with temperature, compared to the experimental study from Yoon et al. [11], and predict that CTE is increasing with a rising temperature over 125 K–600 K. MLMD model also observes a negative to positive transition of CTE at 375 K. Above 600 K, CTE stabilizes and does not increase up-til 1000 K. Classical MD simulations by Gao et al. [41] suggest that thermal fluctuations and out of plane rippling at moderate temperatures can cause the graphene membrane to shrink and exhibit a negative coefficient of thermal expansion. This behavior might not be explicitly visible in DFT simulations where too few atoms are used to observe the rippling. However, classic MD simulations using empirical potentials may not have reliable predictions on the lattice parameter of graphene as shown previously in Fig. 8(a). As for experimental studies, they have the limitations to conduct testing of freestanding graphene sheets. Therefore, our MLMD simulations can bridge this gap and enable classical MD simulations (using MLP) to achieve accuracy comparable to DFT simulations even in larger atomic structures. In our MLMD simulations we have used high sampling of data at moderate temperature range (125 K–500 K), i.e obtained values of CTE at every 25 K interval to accurately capture the thermal rippling and out-of-plane fluctuations. As well as our model has much bigger graphene sheet of size (18 nm *x* 18 nm) which can accurately capture the rippling effect. Thus in moderate temperature range, MLMD simulations predict that CTE of graphene should be much more negative compared to the studies of Mounet et al. [15] and in line with experimental studies of Yoon et al. [11]. In high temperature range (500–1000 K) our results predicts positive CTEs which is consistent with the NGF studies of Jiang et al. [39].

The CTE of graphene is considered to be governed by two competing mechanisms, first is rippling effect which causes CTE to linearly decrease with increase in temperature while the second mechanism is thermal strain which initially causes CTE to non-linearly decrease with temperature (0 K–300 K) and then non-linearly increase with increase in temperature (300 K–1000 K) [41]. These two mechanisms results in an increasing CTE that is negative in temperature range of 0 K–300 K. In this temperature range rate of decrease of thermal strain is decreasing with increasing temperature resulting in a decreasing rate of increase in CTE. Meanwhile, rippling effects has a linearly decreasing impact on CTE in temperature range of 0 K–1000 K. At ∼ 300 K an inflection point is observed in thermal strain [41] curve with respect to temperature. Beyond 300 K thermal strain starts increasing with increase in temperature and nullifies the decreasing effect of thermal rippling in CTE. Thus, CTE switches sign from negative to positive at 375 K. In temperature range of 375 K–1000 K an increasing thermal strain (with reducing slope which tends to become linear beyond 600 K) and linearly decreasing CTE due to rippling effect causes CTE to stabilize and achieve a constant value. This behavior of CTE is accurately captured by MLP as shown in Fig. 8(b).

## 4.2. Young's modulus and ultimate tensile strength

Uniaxial tensile test is one of best methods to validate mechanical properties of graphene and effectiveness of MLP. Computation of stress strain curve was done by loading graphene in armchair direction with a strain rate of 1% per ns between temperature range of 300 K to 1800 K as shown in Fig. 9(a). MLMD simulations were conducted using NPT ensemble with Nosé–Hoover thermostat with a constant pressure set at 1 bar. For each temperature at which Young's modulus and ultimate tensile strength needs to be evaluated simulations were equilibrated for 10 ns before conducting uniaxial tensile test. For each temperature three tensile tests were conducted and Young's modulus and ultimate tensile strength were evaluated by taking average of three similar simulations. As shown in Fig. 9(a) it can be observed that stress strain curve shows a monotonically increasing stress response with increasing strain. Stress strain curve do not show a specific yield point and rate of increase in stress decreases by increasing strain. At higher temperatures a decrease in ultimate tensile strength, decrease in strain to failure as well as decrease in maximum stress can be noticed.

Graphene exhibits a very high elastic modulus (Young's modulus) which makes it one of the stiffest naturally occurring material. Secant modulus was used to evaluate Young's modulus for graphene in this study. Experimental studies at 300 K by Lee et al. [8] predicted a Young's modulus of 1000 ± 100 GPa. DFT-QHA [14] studies also reported a constant Young's modulus of approximately 1044 GPa for graphene in temperature range of 10 K to 1000 K. MD simulations using empirical potentials by Shen et al. [42] predicts a Young's modulus of 1002 GPa in temperature range of 300 K and 700 K. Our MLMD simulations predicted a constant Young's modulus of 975 ± 80 GPa in the temperature range of 300 K to 1800 K suggesting a close match with previously conducted experimental, DFT and MD studies, as shown in Fig. 9(b). The stability of Young's modulus with temperature indicates that graphene maintains its excellent mechanical properties even at higher temperatures and can prove to be an excellent candidate for aerospace applications where superior Young's modulus is a desirable property at higher temperatures.

Ultimate tensile strength is the maximum stress which a material can withstand while being stretched just before breaking. Ultimate tensile strength for graphene is compared with experimental, DFT and MD simulations in Fig. 9(c). Experimental results by Lee et al. [8] at 300 K reported a ultimate tensile strength of 129.7 GPa. DFT-QHA simulations by Tianjiao et al. [14] reported a linearly decreasing ultimate tensile strength with temperature i.e. 119.9 GPa at 50 K and 114.7 GPa at 1000 K. MD studies by Zhao et al. [43] also suggested a linearly decreasing ultimate tensile strength with 93.2 GPa at 300 K and 77.1 GPa at 900 K. Softening in ultimate strength in graphene is caused by weaker inter-atomic interactions due to stronger atomic vibrations with increase in temperature. Our MLMD simulations also exhibits a linearly decreasing ultimate tensile strength with 104.3 GPa at 300 K and 84.85 GPa at 1800 K. MLMD simulation results for ultimate tensile strength are in close match with experimental, DFT and MD studies with a similar trend and thereby validates the effectiveness of MLMD simulations.

## 5. Summary and conclusion

Machine learning potential opens a new direction of research i.e. automating the development of high fidelity interatomic potentials for atomistic simulations to characterize materials with a fraction of cost relative to experimental testing. In this work, we have developed machine learning potentials using artificial neural network and conducted MD simulations using these potentials. Potential energy in these MLMD simulations is evaluated by interpolating the potential energy surface of graphene derived from ANN. In this study, training dataset for DFT simulations was generated by stretching graphene in armchair and zigzag directions. Radial and angular symmetry functions were adopted to represent the atomic environment of reference structures for ANN training. Two types of training methods, GD method and LM method were used in the current work. Comparing with GD method, LM method exhibited great computational efficiency to achieve the target accuracy of the developed MLP.

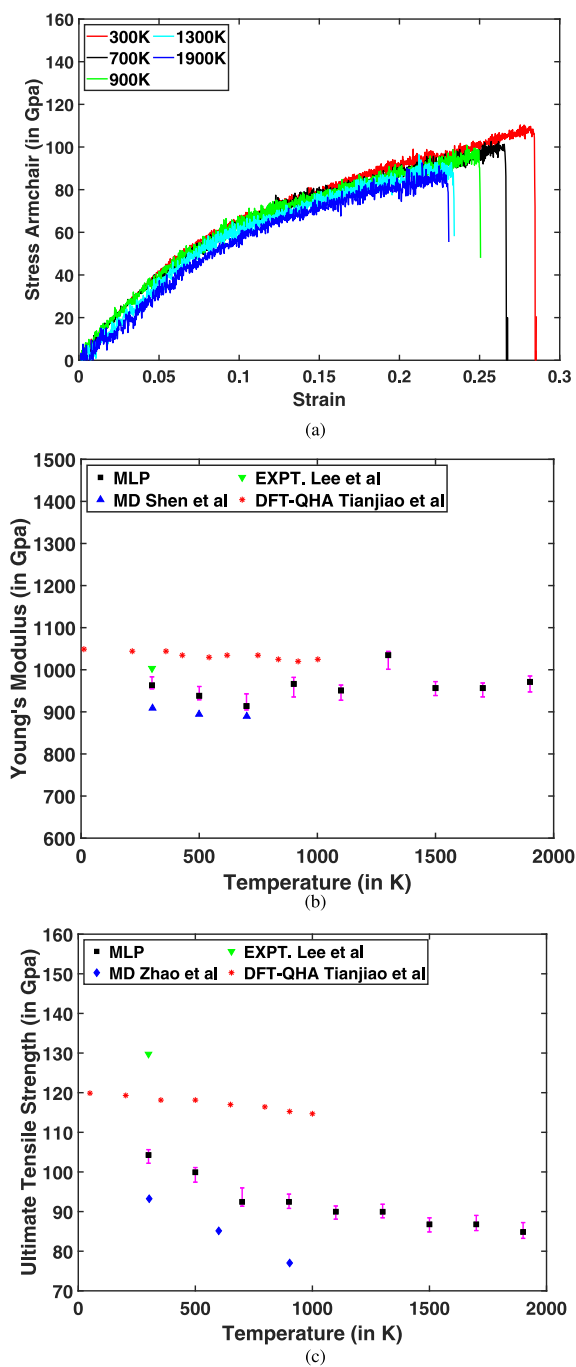

**Fig. 9.** (a) Stress Strain curves for single layer graphene in armchair direction using MLMD simulations for temperature range of 300 K- 2000 K. It can be noticed that the slope of all curves at all temperatures are constant but a decrease in ultimate tensile strength with temperature can be noticed (b) Temperature independence of Young's modulus and comparison with other theoretical and experimental results. (c) Monotonically decreasing ultimate strength found using MLMD simulation and comparison with other theoretical and experimental results. Variation in the evaluated Young's modulus (b) and Ultimate Tensile strength (c) using three identical MLMD simulations can be observed using error bars.

We conducted extensive validation on the quality of the developed MLP. It is demonstrated that a reliable interatomic potential for graphene can be implemented using ANN and symmetry functions to provide accurate predictions about lattice parameter, Young's modulus, ultimate tensile strength and coefficient of thermal expansion for graphene. It was found that both the sampling size of reference training dataset (which can be equivalent to the structural representative of the phase space) as well as the atomic environment representation strategy can have large impact on the accuracy of developed MLP. It is demonstrated that the fidelity of ANN based MLP can be rationally and systematically improved through the control of these determining factors. Furthermore, it is noted that our MLMD simulations can capture the shifting of the governing mechanism of the CTE of graphene over a wide range of temperature between 125 K to 1000 K. This study demonstrates the feasibility of employing ANN to automate the development of interatomic potential for graphene, which can greatly accelerate the discovery and design of innovative graphene based functional materials. Further studies will be conducted to develop MLP for graphene with structural motifs which will enable the characterization of unique material behaviors of graphene.

### CRediT authorship contribution statement

**Akash Singh:** Methodology, Data curation, Validation, Writing. **Yumeng Li:** Conceptualization, Methodology, Writing and review.

### Declaration of competing interest

The authors declare the following financial interests/personal relationships which may be considered as potential competing interests: Yumeng Li reports financial support was provided by University of Illinois Urbana-Champaign.

### Data availability

Data will be made available on request.

### Acknowledgments

The authors acknowledge the support provided by University of Illinois at Urbana Champaign.

### References

[1] D. Zhou, M. Fuentes-Cabrera, A. Singh, R.R. Unocic, J.M.Y. Carrillo, K. Xiao, Y. Li, B. Li, Atomic edge-guided polyethylene crystallization on monolayer two-dimensional materials, Macromolecules 55 (2) (2022) 559–567, http://dx.doi.org/10.1021/acs.macromol.1c01978.

[2] A. Singh, Y. Li, 2D materials guided self-assembly of polymer: molecular dynamics simulation study, in: AIAA SCITECH 2023 Forum, http://dx.doi.org/10.2514/6.2023-0142.

[3] K.S. Novoselov, A.K. Geim, S.V. Morozov, D. Jiang, Y. Zhang, S.V. Dubonos, I.V. Grigorieva, A.A. Firsov, Electric field effect in atomically thin carbon films, Science 306 (5696) (2004) 666–669, http://dx.doi.org/10.1126/science.1102896.

[4] A. Singh, Y. Li, Guided self-assembly of polyethene on graphene, in: AIAA SCITECH 2022 Forum, http://dx.doi.org/10.2514/6.2022-2143.

[5] A.F. Carvalho, B. Kulyk, A.J.S. Fernandes, E. Fortunato, F.M. Costa, A review on the applications of graphene in mechanical transduction, Adv. Mater. n/a (n/a) 2101326, http://dx.doi.org/10.1002/adma.202101326.

[6] A.A. Balandin, S. Ghosh, W. Bao, I. Calizo, D. Teweldebrhan, F. Miao, C.N. Lau, Superior thermal conductivity of single-layer graphene, Nano Lett. 8 (3) (2008) 902–907, http://dx.doi.org/10.1021/nl0731872.

[7] H. Chen, M.B. Müller, K.J. Gilmore, G.G. Wallace, D. Li, Mechanically strong, electrically conductive, and biocompatible graphene paper, Adv. Mater. 20 (18) (2008) 3557–3561, http://dx.doi.org/10.1002/adma.200800757.

[8] C. Lee, X. Wei, J.W. Kysar, J. Hone, Measurement of the elastic properties and intrinsic strength of monolayer graphene, Science 321 (5887) (2008) 385–388, http://dx.doi.org/10.1126/science.1157996.

[9] J.-U. Lee, D. Yoon, H. Cheong, Estimation of Young's modulus of graphene by Raman spectroscopy, Nano Lett. 12 (9) (2012) 4444–4448.

[10] W. Bao, F. Miao, Z. Chen, H. Zhang, W. Jang, C. Dames, C.N. Lau, Controlled ripple texturing of suspended graphene and ultrathin graphite membranes, Nature Nanotechnol. 4 (9) (2009) 562–566, http://dx.doi.org/10.1038/nnano.2009.191.

[11] D. Yoon, Y.-W. Son, H. Cheong, Negative thermal expansion coefficient of graphene measured by raman spectroscopy, Nano Lett. 11 (8) (2011) 3227–3231, http://dx.doi.org/10.1021/nl201488g, pMID: 21728349.

[12] G. Yang, L. Li, W.B. Lee, M.C. Ng, Structure of graphene and its disorders: a review, Sci. Technol. Adv. Mater. 19 (1) (2018) 613–648.

[13] S. Manigandan, P. Gunasekar, S. Nithya, G.D. Revanth, A.V.S.C. Anudeep, Experimental analysis of graphene nanocomposite on Kevlar, IOP Conf. Ser.: Mater. Sci. Eng. 225 (2017) 012061, http://dx.doi.org/10.1088/1757-899x/225/1/012061.

[14] T. Shao, B. Wen, R. Melnik, S. Yao, Y. Kawazoe, Y. Tian, Temperature dependent elastic constants and ultimate strength of graphene and graphyne, J. Chem. Phys. 137 (19) (2012) 194901, http://dx.doi.org/10.1063/1.4766203.

[15] N. Mounet, N. Marzari, First-principles determination of the structural, vibrational and thermodynamic properties of diamond, graphite, and derivatives, Phys. Rev. B 71 (2005) 205214, http://dx.doi.org/10.1103/PhysRevB.71.205214.

[16] S.A. Hollingsworth, R.O. Dror, Molecular dynamics simulation for all, Neuron 99 (6) (2018) 1129–1143, http://dx.doi.org/10.1016/j.neuron.2018.08.011, 30236283[pmid].

[17] Y. Li, G.D. Seidel, Multiscale modeling of the effects of nanoscale load transfer on the effective elastic properties of unfunctionalized carbon nanotube–polyethylene nanocomposites, Modelling Simul. Mater. Sci. Eng. 22 (2) (2014) 025023, http://dx.doi.org/10.1088/0965-0393/22/2/025023.

[18] Y. Li, G. Seidel, Multiscale modeling of functionalized interface effects on the effective elastic material properties of CNT–polyethylene nanocomposites, Comput. Mater. Sci. 107 (2015) 216–234, http://dx.doi.org/10.1016/j.commatsci.2015.05.006.

[19] M.H. Rahman, S. Mitra, M. Motalab, P. Bose, Investigation on the mechanical properties and fracture phenomenon of silicon doped graphene by molecular dynamics simulation, RSC Adv. 10 (2020) 31318–31332, http://dx.doi.org/10.1039/D0RA06085B.

[20] A. Singh, X. Chen, Y. Li, S. Koric, E. Guleryuz, Development of artificial neural network potential for graphene, in: AIAA Scitech 2020 Forum, http://dx.doi.org/10.2514/6.2020-1861.

[21] M. Wen, E.B. Tadmor, Hybrid neural network potential for multilayer graphene, Phys. Rev. B 100 (2019) 195419, http://dx.doi.org/10.1103/PhysRevB.100.195419.

[22] P. Rowe, G. Csányi, D. Alfè, A. Michaelides, Development of a machine learning potential for graphene, Phys. Rev. B 97 (2018) 054303, http://dx.doi.org/10.1103/PhysRevB.97.054303.

[23] A.P. Thompson, H.M. Aktulga, R. Berger, D.S. Bolintineanu, W.M. Brown, P.S. Crozier, P.J. in 't Veld, A. Kohlmeyer, S.G. Moore, T.D. Nguyen, R. Shan, M.J. Stevens, J. Tranchida, C. Trott, S.J. Plimpton, LAMMPS - a flexible simulation tool for particle-based materials modeling at the atomic, meso, and continuum scales, Comput. Phys. Comm. 271 (2022) 108171, http://dx.doi.org/10.1016/j.cpc.2021.108171.

[24] A. Singraber, J. Behler, C. Dellago, Library-based LAMMPS implementation of high-dimensional neural network potentials, J. Chem. Theory Comput. 15 (3) (2019) 1827–1840, http://dx.doi.org/10.1021/acs.jctc.8b00770, pMID: 30677296.

[25] J. Behler, Atom-centered symmetry functions for constructing high-dimensional neural network potentials, J. Chem. Phys. 134 (7) (2011) 074106, http://dx.doi.org/10.1063/1.3553717.

[26] S. Stankovich, D.A. Dikin, G.H.B. Dommett, K.M. Kohlhaas, E.J. Zimney, E.A. Stach, R.D. Piner, S.T. Nguyen, R.S. Ruoff, Graphene-based composite materials, Nature 442 (7100) (2006) 282–286, http://dx.doi.org/10.1038/nature04969.

[27] G. Bebis, M. Georgiopoulos, Feed-forward neural networks, IEEE Potentials 13 (4) (1994) 27–31, http://dx.doi.org/10.1109/45.329294.

[28] P. Giannozzi, S. Baroni, N. Bonini, M. Calandra, R. Car, C. Cavazzoni, D. Ceresoli, G.L. Chiarotti, M. Cococcioni, I. Dabo, A. Dal Corso, S. de Gironcoli, S. Fabris, G. Fratesi, R. Gebauer, U. Gerstmann, C. Gougoussis, A. Kokalj, M. Lazzeri, L. Martin-Samos, N. Marzari, F. Mauri, R. Mazzarello, S. Paolini, A. Pasquarello, L. Paulatto, C. Sbraccia, S. Scandolo, G. Sclauzero, A.P. Seitsonen, A. Smogunov, P. Umari, R.M. Wentzcovitch, QUANTUM ESPRESSO: a modular and open-source software project for quantum simulations of materials, J. Phys.: Condens. Matter 21 (39) (2009) 395502.

[29] N. Artrith, A. Urban, G. Ceder, Efficient and accurate machine-learning interpolation of atomic energies in compositions with many species, Phys. Rev. B 96 (2017) 014112, http://dx.doi.org/10.1103/PhysRevB.96.014112.

[30] M. Fitzner, G.C. Sosso, S.J. Cox, A. Michaelides, The many faces of heterogeneous ice nucleation: interplay between surface morphology and hydrophobicity, J. Am. Chem. Soc. 137 (42) (2015) 13658–13669, http://dx.doi.org/10.1021/jacs.5b08748, pMID: 26434775.

[31] J.H. Los, A. Fasolino, Intrinsic long-range bond-order potential for carbon: Performance in Monte Carlo simulations of graphitization, Phys. Rev. B 68 (2003) 024107, http://dx.doi.org/10.1103/PhysRevB.68.024107.

[32] S.J. Stuart, A.B. Tutein, J.A. Harrison, A reactive potential for hydrocarbons with intermolecular interactions, J. Chem. Phys. 112 (14) (2000) 6472–6486, http://dx.doi.org/10.1063/1.481208.

[33] C.P. Herrero, R. Ramírez, Structural and thermodynamic properties of diamond: A path-integral Monte Carlo study, Phys. Rev. B 63 (2000) 024103, http://dx.doi.org/10.1103/PhysRevB.63.024103.

[34] K.V. Zakharchenko, M.I. Katsnelson, A. Fasolino, Finite temperature lattice properties of graphene beyond the quasiharmonic approximation, Phys. Rev. Lett. 102 (2009) 046808, http://dx.doi.org/10.1103/PhysRevLett.102.046808.

[35] G.A. McQuade, A.S. Plaut, A. Usher, J. Martin, The thermal expansion coefficient of monolayer, bilayer, and trilayer graphene derived from the strain induced by cooling to cryogenic temperatures, Appl. Phys. Lett. 118 (20) (2021) 203101, http://dx.doi.org/10.1063/5.0035391.

[36] Q. Feng, D. Wei, Y. Su, Z. Zhou, F. Wang, C. Tian, Study of thermal expansion coefficient of graphene via raman micro-spectroscopy: revisited, Small 17 (12) (2021) e2006146.

[37] V. Singh, S. Sengupta, H.S. Solanki, R. Dhall, A. Allain, S. Dhara, P. Pant, M.M. Deshmukh, Probing thermal expansion of graphene and modal dispersion at low-temperature using graphene nanoelectromechanical systems resonators, Nanotechnology 21 (16) (2010) 165204, http://dx.doi.org/10.1088/0957-4484/21/16/165204.

[38] S. Linas, Y. Magnin, B. Poinsot, O. Boisron, G.D. Förster, V. Martinez, R. Fulcrand, F. Tournus, V. Dupuis, F. Rabilloud, L. Bardotti, Z. Han, D. Kalita, V. Bouchiat, F. Calvo, Interplay between Raman shift and thermal expansion in graphene: Temperature-dependent measurements and analysis of substrate corrections, Phys. Rev. B 91 (2015) 075426, http://dx.doi.org/10.1103/PhysRevB.91.075426.

[39] J.-W. Jiang, J.-S. Wang, B. Li, Thermal expansion in single-walled carbon nanotubes and graphene: Nonequilibrium Green's function approach, Phys. Rev. B 80 (2009) 205429, http://dx.doi.org/10.1103/PhysRevB.80.205429.

[40] Y. Magnin, G.D. Förster, F. Rabilloud, F. Calvo, A. Zappelli, C. Bichara, Thermal expansion of free-standing graphene: benchmarking semi-empirical potentials, J. Phys.: Condens. Matter 26 (18) (2014) 185401, http://dx.doi.org/10.1088/0953-8984/26/18/185401.

[41] W. Gao, R. Huang, Thermomechanics of monolayer graphene: Rippling, thermal expansion and elasticity, J. Mech. Phys. Solids 66 (2014) 42–58, http://dx.doi.org/10.1016/j.jmps.2014.01.011.

[42] Temperature-dependent elastic properties of single layer graphene sheets, Mater. Des. 31 (9) (2010) 4445–4449.

[43] H. Zhao, N.R. Aluru, Temperature and strain-rate dependent fracture strength of graphene, J. Appl. Phys. 108 (6) (2010) 064321, http://dx.doi.org/10.1063/1.3488620.